\documentclass[10pt,conference]{IEEEtran}

\usepackage{subcaption}

\usepackage{algorithm}
\usepackage{algpseudocode}
\usepackage{amsmath}
\usepackage{amsfonts}
\usepackage{amssymb}
\usepackage{amsthm}
\usepackage{balance}
\usepackage{graphicx}
\usepackage{xspace}
\usepackage{pifont}
\usepackage{multirow}
\usepackage{array}
\usepackage{booktabs}
\usepackage{microtype}
\usepackage{color,colortbl}
\usepackage{dblfloatfix}
\usepackage{framed}
\usepackage[sort]{cite}
\usepackage{xtab,afterpage}
\usepackage{tcolorbox}

\begingroup\expandafter\expandafter\expandafter\endgroup
\expandafter\ifx\csname IncludeInRelease\endcsname\relax
  \usepackage{fixltx2e}
\fi

\usepackage{listings}

\newcommand\lt[1]{{\lstinline+#1+}} 
\renewcommand\t[1]{{\lstinline+#1+}} 
\definecolor{dkgreen}{rgb}{0,0.5,0}
\definecolor{dkred}{rgb}{0.5,0,0}
\definecolor{gray}{rgb}{0.5,0.5,0.5}
\let\origthelstnumber\thelstnumber
\makeatletter
\newcommand*\Suppressnumber{%
  \lst@AddToHook{OnNewLine}{%
    \let\thelstnumber\relax%
     \advance\c@lstnumber-\@ne\relax%
    }%
}

\newcommand*\Reactivatenumber{%
  \lst@AddToHook{OnNewLine}{%
   \let\thelstnumber\origthelstnumber%
   \advance\c@lstnumber\@ne\relax}%
}

\definecolor{LightGray}{gray}{0.9}
\definecolor{Gray}{gray}{0.8}

\usepackage{tikz}

\usepackage{hyperref}
\hypersetup{%
  pdftitle = {title},
  pdfkeywords = {},
  pdfauthor = {Manish Motwani, Ryley Taketa, Larry Golding, 
  Harleen K. Kohli, Yuriy Brun, and Michael C. Fanning},
  bookmarksnumbered,
  bookmarksopen=false,
  urlcolor=black,
  linkcolor=black,
  citecolor=black,
  colorlinks=true,
  pdfstartview={FitH},
}

\def\BibTeX{{\rm B\kern-.05em{\sc i\kern-.025em b}\kern-.08em
    T\kern-.1667em\lower.7ex\hbox{E}\kern-.125emX}}
\begin{document}

\newcommand{\chisq}{\star}
\newcommand{\mannw}{\oplus}
\newcommand{\sysname}{RAFL\xspace}
\newcommand{\sbir}{SBIR\xspace}
\newcommand{\irflname}{Blues\xspace}
\newcommand{\gandv}{G\&V\xspace}
\newcommand{\rfname}{JaRFly\xspace}
\newcommand{\techname}{Swami\xspace}

\title{High-Quality Automated Program Repair}

\author{
Manish Motwani\\
College of Information and Computer Sciences \\
University of Massachusetts Amherst, USA \\
mmotwani@cs.umass.edu
}
\maketitle

\thispagestyle{plain}
\pagestyle{plain}

\begin{abstract}
Automatic program repair~(APR) has recently gained attention because it 
proposes to fix software defects with no human intervention. 
To automatically fix defects, most APR tools use 
the developer-written tests to (a)~localize the defect, 
and (b)~generate and validate the automatically produced 
candidate patches based on the constraints imposed by the tests.
While APR tools can produce patches that appear to fix the defect 
for 11--19\% of the defects in real-world software, 
most of the patches produced are not \emph{correct} 
or acceptable to developers because they overfit to the tests 
used during the repair process. 
This problem is known as the \emph{patch overfitting} problem. 
To address this problem, I propose to equip APR tools 
with additional constraints derived from natural-language software artifacts 
such as bug reports and requirements specifications that 
describe the bug and intended software behavior but are not typically 
used by the APR tools. 
I hypothesize that patches produced by APR tools while using such additional 
constraints would be of higher quality. 
To test this hypothesis, I propose an automated and objective approach 
to evaluate the quality of patches, and propose two novel methods 
to improve the fault localization and developer-written test suites using 
natural-language software artifacts. 
Finally, I propose to use my patch evaluation methodology 
to analyze the effect of the improved fault localization 
and test suites on the quality of patches produced by APR tools
for real-world defects. 
\end{abstract}

\section{Research Problem and Hypothesis}
\label{sec:introduction}

Automated program repair~(APR) tools aim to reduce the cost 
of manually fixing bugs by automatically producing patches~\cite{Gazzola19, Goues19, Monperrus18}.
APR tools have been successful enough to be used in industry~\cite{Bader19, Marginean19}.
The goal of APR tools is to take a program and a suite of tests, 
some of which that program passes and some of which it fails, 
and to produce a patch that makes the program pass all the tests in that suite. 
Unfortunately, these patches can repair some functionality encoded by the tests, 
while simultaneously breaking other, undertested functionality~\cite{Smith15fse}. 
Thus, \emph{quality} of the resulting patches is a critical concern. 
Recent results suggest that the \emph{patch~overfitting} problem\,---\,patches 
pass a particular set of test cases supplied 
to the APR tool but fail to generalize to the desired specification\,---\, is 
common~\cite{Smith15fse, Qi15, Long16a, Le18} and more than 50\% of the 
patches produced by APR tools overfit to the tests used in the repair process~\cite{Motwani20}. 
This makes developers lose trust in the APR tools deterring their wide-scale adoption 
in practice~\cite{Alarcon20}.
The goal of my dissertation is to improve the quality of APR tools. 

Most state-of-the-art APR tools use developer-written tests 
to localize the defect, and generate and validate the automatically 
produced candidate patches based on the constraints imposed by the tests.
While test suites provide an easy-to-use (because they are executable) 
specification, software typically contains many more artifacts that 
describe the desired \emph{correct} software behavior. 
Many of these software artifacts such as requirements specifications, code comments, 
and bug reports use natural-language text to describe the bug 
and intended software behavior, and are therefore not directly 
used by the APR tools. 
I hypothesize that \emph{if I can derive executable constraints from 
such natural-language software artifacts and equip APR tools with these 
additional constraints, it could further constraint the search space 
of the candidate patches and would improve the quality of patches produced.} 
The central goal of my dissertation is to test this hypothesis and for that 
I divide my dissertation work into the following four thrusts:
\begin{enumerate}
    \item[1.] \textbf{Evaluating patch quality~(Section~\ref{sec:jarfly}).} 
    The goal of this thrust is to develop an objective and scalable methodology 
    to evaluate the quality of patches produced by APR tools.   
    I plan to use this methodology to evaluate the quality 
    of patches produced by APR tools for the real-world defects, and analyze how 
    test suite characteristics correlate with patch quality. 

    \item[2.] \textbf{Improving fault localization~(Section~\ref{sec:rafl}).} 
    The goal of this thrust is to develop an approach that uses multiple 
    information sources such as bug reports and test suites 
    to locate defective program elements. 
    I plan to evaluate this approach to localize 
    defects in large, real-world programs and study the effect of 
    improved fault localization on patch quality. 

    \item[3.] \textbf{Improving test suites~(Section~\ref{sec:swami}).}  
    The goal of this thrust is to develop a technique to generate
    executable tests with oracles from natural language software specifications.   
    I plan to evaluate this technique by generating tests from 
    the publicly accessible and reliable specifications of 
    real-world software and analyze the effectiveness of 
    the generated tests. Further, I plan to study the effect of 
    improved test suites on patch quality. 
    
    \item[4.] \textbf{Investigating the effect of improved 
    fault localization and test suites on patch quality~(Section~\ref{sec:part4}).}
    The goal of this thrust is to put together the improved 
    fault localization~(Section~\ref{sec:rafl})
    and developer-written test suites~(Section~\ref{sec:swami}), 
    and then to use the proposed patch evaluation methodology~(Section~\ref{sec:jarfly})
    to validate the proposed hypothesis using real-world defects. 
    
\end{enumerate}

APR tools typically follow a three-step process: 
(1)~identify the location of the defect~(\emph{fault localization}), 
(2)~generate candidate patches by modifying defective location~(\emph{patch generation}), 
and (3)~validate if the candidate patch fixes the defect~(\emph{patch validation}).
The method used for each of these steps can significantly affect the tool's success.
Existing research in APR has mostly focused on devising novel patch generation
algorithms (e.g., heuristic-based~\cite{LeGoues12b, Long16, Tian17, Wen18, Jiang18}, 
constraint-based~\cite{Afzal19, Wang18, Gulwani18, Mechtaev18}, 
and learning-based~\cite{Chen19, Gupta17, Saha17}) aimed to 
produce more correct patches. 
Recently, researchers have started investigating the effect of using 
different technologies, assumptions, and adaptations of 
fault localization techniques~\cite{Assiri17, Yang18, Sun18, Koyuncu19, Jiang19, Lou20}, 
and patch validation methodologies~\cite{Yang17fse, Xiong18, Yu19, Gao19, Tian20, Wang20} 
on the performance of APR tools. 
Recent patch validation methodologies employ machine learning 
models to identify overfitted patches based on their source code features.
Unlike existing methods, my proposed methods aim to improve the steps 
of the repair process by using information derived from natural-language software artifacts. 
The improved fault localization shall enable APR tools to produce patches 
for the correct defective locations while the improved test suites will increase 
the constraints imposed on candidate patches during the patch generation and validation
steps, and therefore shall improve the quality of produced patches.

\section{Research Contributions and Results So Far}
\label{sec:research-contributions}
This section describes the research contributions 
organized in terms of the four thrusts of the dissertation. 
I describe each contribution in the context of existing research 
work and present the results obtained so far. 

\subsection{Evaluating Patch Quality}
\label{sec:jarfly}

To address the patch overfitting problem, we first need 
a method to evaluate the quality of the produced patches.
Prior studies of quality of APR have either used manual
inspection~\cite{Qi15, Martinez17}, or have
used automatically generated, independent, evaluation test suites 
not used during the repair process~\cite{Smith15fse, Ye19}.
The issue with manual inspection is that it cannot scale 
to evaluate hundreds of automatically produced patches and 
can be subject to subconscious bias, especially if the inspectors 
are authors of the tools being evaluated~\cite{Le19}. 
Contrastingly, using evaluation tests is inherently partial, 
as the generated tests may undertest the patched program. 
Existing studies that use evaluation tests focus on small programs 
and relatively-easy-to-fix defects~\cite{Smith15fse, Ye19}.

To address these issues, I proposed a methodology~\cite{Motwani20} that uses 
\emph{high--quality} evaluation test suites to evaluate 
the quality of the produced patches. 
My methodology uses today's state-of-the-art 
test-suite generation techniques and overcomes 
their shortcomings to produce high-quality test suites.  
The automatically generated evaluation test suites used 
in my methodology cover 100\% of all the developer-modified methods 
and at least 80\% of all the developer-modified classes. 
My methodology ensures that the evaluation test suites 
do not undertest the patched program. 

I performed a detailed study~\cite{Motwani20} to 
investigate the patch overfitting problem and identify 
the correlation between test suite characteristics and patch quality. 
I evaluated four representative APR tools~(GenProg~\cite{LeGoues12a}, 
TrpAutoRepair~\cite{Qi13}, Par~\cite{Kim13}, and SimFix~\cite{Jiang18}) 
on 357~real-world defects in 5~large, complex Java projects 
from the Defects4J benchmark~\cite{Just14defects4j}. 
My evaluation employed rigorous statistical analyses and controlled 
for confounding factors to increase the likelihood that my results generalize. 
I answered four research questions:

\textbf{RQ1.1~How often and how much do the patches produced by APR
tools overfit to the developer-written test suite and fail to generalize
to the evaluation test suite, and thus ultimately to the program
specification?}
Often. For the four tools I evaluated, only
between 13.8\% and 41.6\% of the patches pass 100\% of an independent test
suite. Patches typically break more functionality than they repair. 

\textbf{RQ1.2~How do the coverage and size of the test suite used to
produce the patch affect patch quality?}
Larger test suites produce slightly higher-quality
patches, though, surprisingly, the effect is extremely small. Also
surprisingly, higher-coverage test suites correlate with lower quality, but,
again, the effect size is extremely small. 

\textbf{RQ1.3~How does the number of tests that a buggy program fails
affect the degree to which the generated patches overfit?}
The number of failing tests correlates with
slightly higher quality patches.  

\textbf{RQ1.4~How does the test suite provenance (whether it is written
by developers or generated automatically) influence patch quality?}
Test suite provenance has a significant effect on
patch quality, although the effect may differ for different APR tools. In
most cases, human-written tests lead to higher-quality patches. \medskip

\begin{tcolorbox}
These results corroborate the patch overfitting problem in the state-of-the-art
APR tools. Further, these results indicate that improving the quality 
of test suites used in the repair process can potentially improve the quality 
of automatically produced patches. 
\end{tcolorbox}

\subsection{Improving Fault Localization}
\label{sec:rafl}
Fault localization~(FL) is recently identified as a key aspect 
of APR that affects patch correctness~\cite{Afzal19, Liu19, Jiang19, Wen17, Assiri17, Yang18}.
To identify the defective program elements, most APR tools 
use spectrum-based fault localization~(SBFL), which uses
test-execution coverage to compute the suspiciousness
scores of the program's elements, such as classes, methods, and statements. 
The elements are ranked based on these scores and APR tools use top-ranked 
elements as candidate locations to patch defects.
To the best of my knowledge, only two APR tools, R2Fix~\cite{Liu13} 
and iFixR~\cite{Koyuncu19}, use information retrieval-based fault 
localization~(IRFL), which ranks suspicious program elements 
based on their similarity with the text in bug reports. 
Using SBFL and IRFL can be complementary. For example, iFixR patches
defects that 16~SBFL APR tools cannot, and vice
versa~\cite{Koyuncu19}.
Recent studies also show that combining FL~techniques that 
use different information sources (e.g., SBFL using test suites 
and IRFL using bug reports) can significantly outperform individual 
FL~techniques in terms of localizing defects~\cite{Zou19, Li19, Le16issta}. 
Based on these findings, I hypothesize that using combined SBFL and IRFL 
can enable APR tools to patch all the defects that they can patch 
when using the underlying SBFL and IRFL alone, and perhaps some others. 
To the best of my knowledge, this is the first investigation of the effect 
of combined FL on APR.

Existing approaches~\cite{Li19, Zou19, Sohn17, Le16issta, Xuan14icsme} 
to combine multiple FL techniques, are based on \emph{learning to rank}~\cite{Burges05}, 
supervised deep machine learning techniques. The performance and generalizability 
of such approaches depend heavily on the dataset and features used 
for training the machine learning model. 
I proposed to use an unsupervised approach that requires no training.
To combine FL~techniques, I developed 
Rank~Aggregagtion-based~FL~(\sysname)~\cite{Motwani20arxiv}, a novel 
approach that uses rank aggregation algorithms~\cite{Lin10} to combine 
the top-k ranked statements produced using different FL~techniques. 
\sysname measures the similarity of the two ranked lists using 
the Spearman footrule distance~\cite{Brandenburg13} and runs 
the cross-entropy Monte Carlo algorithm~\cite{Rubinstein13} to 
produce a super list of top-k statements while maximizing the similarity 
to the individual lists. 
\sysname can combine FL results obtained using any set of techniques; 
in my dissertation, I specifically focus on combining 
SBFL and IRFL, as these two are used in APR.

Existing IRFL techniques~\cite{Zhou12, Saha13, Wong14, Youm15, Wen16} are not
well suited for APR because they localize defects at the file or
method level, whereas APR tools need statement-level granularity. 
I developed \irflname~\cite{Motwani20arxiv}, a statement-level IRFL technique 
based on BLUiR~\cite{Saha13}, an
existing file-level IRFL technique. \irflname considers the abstract syntax
tree~(AST) representations of program statements as a collection of documents,
and bug report as a query, and uses a structured information retrieval
technique to rank the statements based on their similarity with the bug
report.
\irflname is the first unsupervised statement-level IRFL technique. The prior
statement-level IRFL technique, D\&C~\cite{Koyuncu19arxiv} used by
iFixR~\cite{Koyuncu19}, requires supervised training.

I implemented an SBFL technique using GZoltar~v1.7.2,
and the Ochiai ranking strategy, which is one of the most effective ranking
strategies in object-oriented programs~\cite{Xuan14icsme, Zou19}. 
I evaluated this SBFL technique, \irflname, and their \sysname-enabled
combination \sbir, on 818 real-world defects from 17 large Java projects 
in the Defects4J~v2.0 benchmark~\cite{GayJ2020}.

To test if the combined FL improves the quality of patches,  
I used SimFix~\cite{Jiang18}, a state-of-the-art APR tool. 
I chose SimFix because a recent study~\cite{Koyuncu19} found that it 
outperforms a suite of 16 other APR tools, including 
iFixR, kPAR~\cite{Liu19}, AVATAR~\cite{Liu19saner}, and LSRepair~\cite{Liu18}, 
as well as others.
I ran SimFix on 818~defects in Defects4J~v2.0 for which bug reports 
are available using my SBFL implementation, \irflname, and \sbir.
To evaluate the correctness of patches, I used my patch evaluation 
methodology~(recall Section~\ref{sec:jarfly}).
I answered three research questions: 

\textbf{RQ2.1~Does SBIR localize more defects 
than the underlying SBFL and Blues techniques?}
Yes, \sbir outperforms SBFL and \irflname, for
all sizes of suspicious statement lists investigated (1, 25, 50, 100). For
example, \sbir correctly identifies a buggy statement as the most
suspicious for 148 of the 818 (18.1\%) defects, whereas SBFL does so
for 89~(10.9\%) and \irflname for 25~(3.1\%). 

\textbf{RQ2.2~Does using SBIR in APR patch more defects?}
Using \sbir enables SimFix to patch marginally 
more defects (112 out of 818) than using SBFL (110) and significantly 
more than using \irflname (55). 
With \sbir, SimFix produces patches for most of the defects patched
using SBFL or \irflname, as well as 3 new defects that could not be patched
previously. Further, with \sbir, SimFix can produce all but one (29
out of 30) of the correct patches it produces with SBFL, and all (16 out of
16) of the correct patches with \irflname.
Additionally, SimFix with my FL implementations patches 
10 defects that none of 14 state-of-the-art APR tools patch~\cite{Liu19}.
Finally, using \irflname, SimFix significantly outperforms iFixR, the
state-of-the-art IRFL-based APR tool~\cite{Koyuncu19}, patching 19 out of
156~defects (7 correctly) while iFixR patches only 4 defects (3 correctly). 

\textbf{RQ2.3~How does the patch quality vary across the
new and old versions of Defects4J benchmark?}
Past APR evaluations fail to generalize to new defects.
For example, SimFix correctly patches 3--6\% (6\% when using SBFL,
3\%~\irflname, 6\%~\sbir) of the defects in the older version of
Defects4J, but only 1--2\% (2\%~SBFL, 1\%~\irflname, 2\%~\sbir) of the
\emph{new} defects. Of the patches SimFix produces for
the old defects, 39--40\% are correct; for the new defects, only 13--19\%
are correct. \medskip

\begin{tcolorbox}
These results show that improving FL 
can enable APR tools to patch new defects without requiring any 
changes to their core patch generation and validation algorithms.
A recent study~\cite{Liu20} shows that the quality of patches
produced by some APR tools is more sensitive to the accuracy of 
FL~results they use than others.
Hence, I plan to extend my evaluation of using improved fault localization
to more sensitive APR tools. 
\end{tcolorbox}

\subsection{Improving Test Suites}
\label{sec:swami}

The developer-written tests are often inadequate~\cite{Lawrence05}
yet they are used by most APR tools because the tests are
readily available and are machine-processable. 
Tests consist of two parts, an input to trigger a behavior 
and an oracle that indicates the expected behavior.
While the state-of-the-art automated test generation 
techniques~(e.g., Randoop~\cite{Pacheco07a}, EvoSuite~\cite{Fraser13})
can effectively generate test inputs, they require
a reference implementation to compute oracles for the generated inputs. 
In practice, a correct reference implementation 
may not be available, thus, limiting the use of such 
test generation techniques to improve the quality of APR tools. 
To address this, I analyzed other software artifacts from which I
can derive the intended software behavior and improve 
the developer-written tests, which are used by APR tools.
While formal, mathematical specifications that can be used
automatically by computers are rare, developers do write natural language~(NL)
specifications, often structured (e.g., JavaDoc comments), 
as part of software requirements specification documents. 
Hence, in this thrust, I tackle the problem 
of automatically generating tests from such structured NL 
specifications to verify that the software does what the specifications say it should. 

I developed \techname~\cite{Motwani19icse}, a technique to automatically generate 
executable tests from structured NL specifications. 
I scoped my work by focusing on exceptional and
boundary behavior, precisely the important-in-the-field behavior developers
often undertest~\cite{Goffi16, Weimer04}.
\techname uses regular expressions to identify what sections of the specification
document encode testable behavior.
\techname then applies a series of four regular-expression-based rules to
extract information about the syntax for the methods to be tested, the
relevant variable assignments, and the conditionals that lead to visible
oracle behavior, such as return statements or exception throwing statements.
\techname then backtracks from the visible-behavior statements to recursively
fill in the variable value assignments according to the specification,
resulting in a test template encoding the oracle, parameterized by test
inputs. \techname then generates random, heuristic-driven test inputs to
produce executable tests.

\techname complements prior work~(e.g, ~\cite{Fraser13,Pacheco07a}) 
on automatically generating test inputs for regression tests or manually-written oracles
by automatically extracting oracles from NL specifications. The closest work to
\techname is Toradacu~\cite{Goffi16} and Jdoctor~\cite{Blasi18}, which focus on
extracting oracles for exceptional behavior, and @tComment~\cite{Tan12},
which focuses on extracting preconditions related to nullness of parameters.
These techniques are limited to using Javadoc comments, which are simpler
than the specifications \techname tackles. 
\techname builds on these techniques, expanding the rule-based 
NL processing techniques to apply to more complex NL.
Additionally, unlike those techniques, \techname generates oracles for 
boundary conditions along with exceptional behavior. 

I evaluated \techname using ECMA-262, the official specification of the
JavaScript programming language~\cite{Wirfs-Brock17}, and two well known
JavaScript implementations: Java Rhino and C++ Node.js. 
I answered three research questions:

\textbf{RQ3.1~How precise are \techname-generated tests?} Of the tests
\techname generates, 60.3\% are innocuous\,---\,they can never fail. Of the
remaining tests, 98.4\% are precise to the specification and only 1.6\% might 
raise false alarms.

\textbf{RQ3.2~Do \techname-generated tests cover behavior missed by
developer-written tests?} \techname-generated tests identified 1 previously
unknown defect and 15 missing JavaScript features in Rhino, 1 previously
unknown defect in Node.js, and 18 semantic ambiguities in the ECMA-262 specification.
Further, \techname generated tests for behavior
uncovered by developer-written tests for 12 Rhino methods. The average
statement coverage for these methods improved by 15.2\% and the average branch
coverage improved by 19.3\%. 

\textbf{RQ3.3~Do \techname-generated tests cover behavior missed by
state-of-the-art automated test generation tools?} Comparing \techname 
tests to EvoSuite tests revealed that most of the EvoSuite tests that
cover exceptional behavior were false alarms, whereas 98.4\% of the
\techname tests were precise to the specification and can only
result in true alarms. Augmenting EvoSuite tests using \techname
increased the statement coverage of 47 classes by, on average, 19.5\%.
\techname also produced fewer false alarms than Toradacu
and Jdoctor, and, unlike those tools, generated tests 
for missing features. 

\begin{tcolorbox}
To study the effect of improved test suites on patch quality, 
I plan to create a dataset of defects for which \techname can
be used to improve developer-written tests.  
I will then perform controlled experiments on that dataset using APR tools 
to patch those defects by using the original and developer-written tests
augmented with \techname tests in the repair process. 
\end{tcolorbox}

\subsection{Investigating the effect of improved 
fault localization and test suites on patch quality.}
\label{sec:part4}
This thrust aims to put together the improved fault localization~(FL)~(Section~\ref{sec:rafl}) 
and test suites~(Section~\ref{sec:swami}), and then use 
the proposed patch evaluation methodology~(Section~\ref{sec:jarfly}) to measure 
the repair success.
I will first create a defect dataset on which I can run different FL 
techniques, as well as improve test suites using Swami. 
Next, I will select a representative set of APR tools that are more sensitive 
to FL accuracy and use them to patch the defects. 
Finally, I will study the individual as well as 
the combined effect of improving FL and test suites on the patch quality. 
I will answer three research questions:

\textbf{RQ4.1~Does improving fault localization using \sbir improves patch quality?}

\textbf{RQ4.2~Does improving test suites using \techname improves patch quality?} 

\textbf{RQ4.3~Does improving both fault localization and test suites improve patch quality?}

\section{Expected Contributions and Timeline}
\label{sec:research plan}

My dissertation will make 
the following contributions that will extend the state-of-the-art 
of APR along with other closely related fields. 
I plan to release all of the data and tools developed as part of my dissertation 
as freely available and reusable open-source software.

\begin{enumerate}
    \item An APR quality evaluation framework that provides 
    an automated and objective methodology to evaluate patch quality
    and allows APR tools to use multiple fault localization~(FL) techniques in the repair process. 
    \item \rfname\footnote{\url{http://jarfly.cs.umass.edu}}, the Java Repair Framework, 
    which simplifies the implementation of Java techniques for genetic improvement 
    including but not limited to genetic improvement techniques for APR.
    \item \sysname, an unsupervised learning-based approach 
    to improve the FL accuracy by combining results of multiple
    FL techniques that use different information sources.
    \item \irflname, the first unsupervised learning-based statement-level IRFL technique 
    that can be used by the APR tools to localize and patch defects using bug reports.
    \item \techname\footnote{\url{http://swami.cs.umass.edu}}, a novel approach 
    to generate tests with oracles from structured natural language software 
    specifications for exceptional behavior and boundary conditions.
\end{enumerate}

I am currently working on the fourth thrust~(Section~\ref{sec:part4})
and I plan to finish all of my dissertation work by late 2021.  

\balance

\bibliographystyle{IEEEtran}
\bibliography{IEEEabrv,relatedwork}

\end{document}